\documentstyle[prl,aps,graphicx,multicol]{revtex}
\input psfig.sty

\begin{document}
\title{\bf On the origin of heavy quasiparticles in LiV$_2$O$_4$}
\vspace{1.0em}
\author{P.~Fulde$^a$, A.~N.~Yaresko$^{b,c}$, A.~A.~Zvyagin$^{a,d}$ and
Y.~Grin$^b$}  
\address{$^{a}$ Max Planck Institut f\"ur Physik komplexer Systeme,
N\"othnitzer Strasse, 38, D-01187 Dresden, Germany}
\address{$^{b}$ Max Planck Institut f\"ur Chemische Physik fester Stoffe,
N\"othnitzer Strasse, 40, D-01187 Dresden, Germany}
\address{$^{c}$ Institute for Metal Physics of the National Academy of 
Sciences of Ukraine, 36 Vernadskii Avenue, Kiev, 01142, Ukraine}
\address{$^{d}$B.~I.~Verkin Institute for Low Temperature Physics and
Engineering, \\ of the National Academy of Sciences of Ukraine, 47 Lenin 
Avenue, Kharkov, 61164, Ukraine}

\date{\today}
\maketitle

\begin{abstract}
An explanation is provided for the heavy quasiparticle excitations in
LiV$_2$O$_4$. It differs considerably from that of other known heavy-fermion
systems. Main ingredients of our theory are the cubic spinel structure of the
material and strong short-range correlations of the d electrons. The large
$\gamma$ coefficient is shown to result from excitations of Heisenberg spin
$\frac{1}{2}$ rings and chains. The required coupling constant is calculated
from LDA+U calculations and is found to be of the right size. Also the
calculated Sommerfeld-Wilson ratio is reasonably close to the observed one.\\  
\end{abstract}

\centerline{PACS numbers: 71.27.+a, 71.10.-w, 75.40.Cx}

\begin{multicols}{2}
\narrowtext
The low-temperature properties of the metal LiV$_2$O$_4$ show many
of the characteristic features of heavy-fermion systems like CeAl$_3$ or
CeRu$_2$Si$_2$ \cite{ref1}. The coefficient $\gamma$ of the low-temperature
specific heat $C = \gamma T$ is large, i.e., $\gamma = 0.35-0.42 J/mol K^2$
\cite{ref1}, depending on samples and $C(T)$ shows a broad maximum around $16
K$. The large $\gamma$ value implies a high density of fermionic low-energy
excitations. In fact, when $C(T)$ is integrated up to $60 K$ it is found that
the associated entropy is close to $2R \ln 2$ where $R$ is the gas
constant. This suggests that there is roughly one excitation per $V$ ion in the
system when it is heated up to that temperature, a finding inconceivable for a
conventional, partially filled conduction band of $d$-electrons. In fact, the
$\gamma$ value obtained from ab initio band structure calculations
\cite{ref4,ref5,ref6,ref7} turns out to be too small by a factor of 25. In
accordance with the observed large $\gamma$ value it is found that also the
spin susceptibility $\chi_{sp}(T)$ is similarly enhanced \cite{ref2}. It is
nearly $T$ independent below $T \approx 30 K$ with a shallow broad maximum
around $16 K$ and a sample dependent increase below $5 K$. The
Sommerfeld-Wilson ratio $R_W = \pi^2 k_B^2 \chi_{sp}/(3 \mu^2_{\rm eff}
\gamma)$ is 1.7 with considerable error bars because of the remaining low
temperature variation of $\chi_{sp}(T)$. A ratio of that size shows that the
enhancements in $\gamma$ and $\chi_{sp}$ nearly cancel each other as expected
for a metal with an enhanced density of states $N^*(0)$ at the Fermi
energy. Remember that $\gamma \sim N^*(0)$ and $\chi_{sp} \sim N^*(0)$ in that
case. No (anti)ferromagnetic order was found above $0.02 K$ and no evidence for
spin-glass behavior. There are also no indications of charge ordering, i.e.,
all $V$ sites are equivalent. The low-temperature resistivity is $\rho(T) =
\rho_0 + AT^2$ with a large coefficient $A = 2\mu \Omega cm/K^2$ like in other
heavy-fermion systems. LiV$_2$O$_4$ is therefore a rare case of a system with
heavy quasiparticle excitations which do not involve $f$ electrons \cite{ref1}.

The physical origin of the low-energy scale associated with the characteristic
heavy-fermion excitations has remained unclear. Presently we know of a few
physical mechanisms resulting in heavy quasiparticles which all seem
inapplicable here. 

One well known source is the Kondo effect \cite{ref8}. It applies to
heavy-fermion systems like CeAl$_3$ and requires ions with quasi-localized $f$
electrons of nearly integer number. In fact $\gamma$ is the larger the closer
the $f$-count $n_f$ is to an integer, e.g., to $n_f = 1$ in the case of Ce
ions. But in LiV$_2$O$_4$ the valency of the $V$ ions is +3.5 and therefore the
Kondo effect is excluded here. A second mechanism is realized in
Nd$_{2-x}$Ce$_x$CuO$_4$ $(x \approx 0.1-0.2)$ and is essentially based on the
Zeeman effect \cite{ref9}. It applies to systems with strongly correlated
conduction electrons which couple weakly to ions with localized spins, i.e., Nd
ions in the above case. This  situation can be also excluded for
LiV$_2$O$_4$. The same holds true for heavy fermions appaering very close to a
quantum critical point, a model which has been suggested for YMn$_2$
\cite{ref10}. There remains the case of heavy quasiparticles caused by charge
ordering like in Yb$_4$As$_3$. Here the characteristic feature is the formation
of well separated spin $\frac{1}{2}$ Heisenberg chains in a three-dimensional 
lattice due to charge order \cite{ref11,ref12}. This route seems promissing
except that there is no charge ordering in LiV$_2$O$_4$. In that material all
$V$ sites remain equivalent down to the lowest temperatures. Nevertheless, it
is this example which leads us to suggest the physical model discussed in the
following. 

The aim of this communication is to discuss a new scenario for the
origin of a low-energy scale. It uses the fact that LiV$_2$O$_4$ has a cubic
spinel structure down to the lowest temperatures measured \cite{ref2,ref13},
i.e., the system seemingly does not charge order. But it has obviously strong
short-ranged electron correlations. The spinel structure enables us to account
for them in a particular way which suggests directly a plausible explanation of
the large number of fermionic low-energy excitations as evidenced by the large
$\gamma$ coefficient. 

In the spinel structure the V ions form corner-sharing tetrahedra. Their
sublattice can be viewed as a fcc lattice with every second site removed. A $V$
ion has six nearest neighbors of its kind and the $V$ ions are the centers of
slightly distorted edge sharing octahedra of oxygen ions. All $V$ ions are
crystallographically equivalent. Spinel structures or pyroclore lattices give
raise to frustrated antiferromagnetic interactions. Resulting forms of a
spin-liquid ground-state have been frequently discussed see, e.g.,
\cite{ref14,ref15,ref16}. 
 
Band structure calculations employing the LDA+$U$
computational method show that the Fermi energy is placed in vanadium $t_{2g}$
bands which are well separated from the oxygen $p$-bands. The average
$d$-electron number per $V$ ion is 1.5. For those electrons we assume the
following simplified model Hamiltonian to apply
\begin{eqnarray}
\label{1} H = & \ & \sum_{\langle ij \rangle, \nu \sigma} t_\nu (a^+_{\nu
\sigma} (i) a_{\nu \sigma} (j) + h.c.) \nonumber \\
&+& U \sum_{i; \nu} n_{\nu \uparrow} (i) n_{\nu \downarrow} (i) + U \sum_{i;
\nu > \mu} n_\nu (i) n_\mu (i)\nonumber \\ 
&+& V_0 \sum_{\langle ij \rangle} n (i) n (j) + J_H \sum_i {\bf s}_1 (i)  {\bf
s}_2 (i)\nonumber \\ 
&+& \sum_{\langle ij \rangle} J_{ij} {\bf S} (i)  {\bf S} (j) ~~~. 
\end{eqnarray}
The $a_{\nu \sigma}^+ (i)(a_{\nu \sigma} (i))$ create (destroy) a $d$-electron
in orbital $\nu$ with spin $\sigma$ at site $i$. For simplicity we shall use
two $d$-orbitals only, one pointing along the octahedron axis and the other
pointing towards the four oxygen ions within the plane. The first term
describes the kinetic energy and only transitions between neighboring sites
$\langle ij \rangle$ are taken into account. The second and third term account
for the on-site Coulomb repulsion of $d$-electrons in the same and in different
orbitals. Here $n_\nu(i) = \sum\limits_\sigma a^+_{\nu \sigma} (i) a_{\nu
\sigma}(i)$ is the number operator. The fourth term describes the Coulomb
repulsion of electrons situated 
on neighboring sites, the fifth term is due to Hund's rule coupling and the
last terms results from superexchange via the oxygen sites. The total spins
${\bf S}_{i(j)}$ of the neighboring sites $i$ and $j$ depend on the number of
$d$ electrons at those sites. For a $d^1$ configuration $S = \frac{1}{2}$ while
for a $d^2$ site $S = 0$ or 1 etc. We assume that $U$ and
$V_0$ are much larger than the $t_\nu$ while $J_{ij}$ is expected to be of
similar size. A large value of $U$ implies that only $d^1$ and $d^2$
configurations need consideration. Moreover, a large value of $V_0$ requires
that on each tetrahedron two sites are in a $d^1 (V^{4+})$ and two are in a
$d^2 (V^{3+})$ configurations (tetrahedron rule). This has been first pointed
out by Verwey \cite{ref17} and was discussed in detail by Anderson
\cite{ref10}. Note, that because of an average $d^{1.5}$ occupancy $d^2-d^2$
and $d^1-d^1$ configurations repel each other while $d^1-d^2$
configurations are attractive. LiV$_2$O$_4$ is a metal and therefore the above
tetrahedron rule cannot be strictly obeyed. Otherwise a single transition $d^2
\rightleftharpoons d^1$ would imply a rearrangement of configurations in
nearly all tetrahedra. However, we are interested here in the spin excitations
only, because of the large entropy associated with the low-energy
excitations. Therefore the heavy quasiparticle excitations must involve
predominantly spin degrees of freedom. Thus for the present purpose we may
assume that the tetrahedron rule is strictly obeyed. In that case the ground
state $| \psi_0 \rangle$ of the system is of the form 
\begin{equation}
\label{2}  \mid \psi_0 \rangle = \sum_n \alpha_n \mid \phi_n \rangle
\end{equation}
where the  $| \phi_n \rangle$ are all configurations which obey the tetrahedron
rule. In each of the $| \phi_n \rangle$ the $V^{4+}$ ions form chains and
rings. They are surrounded by chains and rings of $V^{3+}$ ions. An example is
shown in Fig. 1. One may check that the smallest rings possible involve six
$d^1 (d^2)$ ions. Ring sizes increase in units of four, i.e., they involve
10,14 etc. $V$ ions of one kind. The form of the $\alpha_n$ is irrelevant here
and will be the subject of a separate investigation.

Consider the $d^2$ ions. Without Hund's coupling the ratio of $S = 1$ triplet
and $S = 0$ singlet configurations would be approximately 3 : 1. Finite Hund's
coupling enlarges it further. Therefore we deal with $S = 1$ chains (and rings)
interrupted occasionally by sites with $S = 0$.
\begin{figure}[h]
\centerline{\psfig{file=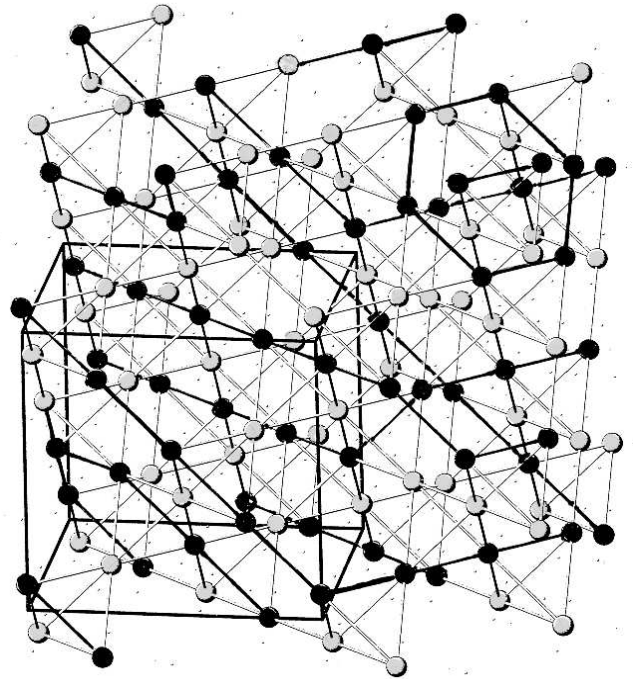,width=8.5cm}}
\caption[l]{Example of a configuration $| \phi_n \rangle$ obeying the
tetrahedron rule. Solid dots: spin $\frac{1}{2}$ sites; empty dots: spin 1
sites. They form chains and rings each. Notice a ring consisting of six spin
$\frac{1}{2}$ sites in the upper right corner.}
\label{fig1}
\end{figure}

Let us assume next that the coupling between $S = 1$ and $S = \frac{1}{2}$
sites would vanish. In that case the excitations of the system are given by
those of the $S = \frac{1}{2}$ chains and rings and of the $S = 1$ chains and
rings, the latter being disrupted by $S = 0$ configurations. At low
temperatures only the excitations of the $S = \frac{1}{2}$ Heisenberg chains
are important because their spectrum goes to zero in the long wave length
limit. The excitations of the $S = 1$ Heisenberg chains and rings have a gap in
the excitation spectrum and play no role at low $T$. The same holds true for $S
= \frac{1}{2}$ rings which are sufficiently small. 

\noindent The relevant excitations are of the form
\begin{equation}
\label{3}  A_k^+ \mid \psi_0 \rangle = \sum_n \alpha_n A_k^+ \mid \phi_n
\rangle 
\end{equation}
where $A_k^+$ generates a spin wave of wavelength $\lambda = \frac{2 \pi}{k}$
in the $S = \frac{1}{2}$ chains and large rings contained in $| \phi_n
\rangle$. The  $S = \frac{1}{2}$ chains and large rings have a much larger
statistical weight than the $S = \frac{1}{2}$ small rings and therefore we may
assume that half of the $N$ sites are involved in the low energy
excitations. The corresponding low-temperature specific heat is $C = \gamma T$
with a $\gamma$ coefficient given by \cite{ref8}
\begin{equation}
\label{4}  \gamma = \frac{2}{3} \frac{k_B R}{J_3}
\end{equation}
where $J_3 = J_{ij}$ when $S(i) = S(j) = \frac{1}{2}$. Note that only one $V$
ion per formula unit is contributing to the low-energy excitations and hence to
$\gamma$. The low-temperature spin susceptibility is that of Heisenberg chains
and given by \cite{ref15} $\chi_{sp} \simeq 4 \mu_{\rm eff}^2 R/(\pi^2 J_3)$
which results in a Sommerfeld-Wilson ratio of $R_W = 2$.

In the following we want to provide arguments which suggests that the above
considerations still apply, when the coupling $J_{ij}$ between sites with spin
$S = \frac{1}{2}$ and $S = 1$ is included. For that purpose we must first
estimate the relative sizes of $J_{ij}$ when spins $\frac{1}{2}$ and spins 1
are coupled among themselves, i.e., $J_3$ and $J_1$ and when a spin
$\frac{1}{2}$ couples with a spin 1, i.e., $J_2$. With this aim we have
performed LDA+$U$ band structure 
calculations in order to estimate their relative sizes. The calculations were
performed under the following assumptions. First of all, we assumed that the
$V$ ions with $d^1$ and $d^2$ configurations form two different families of
ordered chains which alternate in $z$ direction. Both families lie in the $xy$
plane but they differ by a rotation of $\pi/2$. With this arrangement each $V$
tetrahedron is occupied by an equal number of $d^1$ and $d^2$ ions, i.e., the
tetrahedron rule is strictly obeyed. It should be noticed that no charge
ordering of $V$ ions has been observed experimentally. Nevertheless the chosen
configuration serves its purpose. We have found that for such an arrangement of
$V$ ions a self-consistent solution can only be obtained for values of the
screened on-site Coulomb integral $U$ larger than 3.8 eV, where as the use of
smaller values of $U$ leads to a solution with the average $d^{1.5}$
count for each $V$ ion. The value of $U = 4$ eV, that was used in the
present work to calculate $J_{ij}$, is larger than the value of about 3 eV
previously estimated and the experimental one which is close to 2 eV
\cite{ref6,ref20}. However, we believe that the calculated $J_{ij}$
can be used to estimate their relative strengths. In order to calculate the
coupling constants we started from a self-consistent solution for a collinear
magnetic state with an AF ordering of $d^1$ and $d^2$ ions along the
corresponding chains. In this case each $d^1$ ion has two $d^2$ nearest
neighbours with the same and two with the opposite orientation of the spin
direction, i.e., $d^1 - d^2$ magnetic interactions are frustrated. Then, the
band energies of non-collinear magnetic structures with an angle between either
the $d^1$ spins or the $d^2$ spins varying from $180^{\circ}$ to $0^{\circ}$
were calculated, and $J_{ij}$'s were determined by comparing total energies
with those of the corresponding Heisenberg-like model with nearest neighbour
interaction. (For a detailed description of the procedure see Ref. \cite{ref21}
and references therein.) As a result we obtained values of 3 meV and 24 meV for
the coupling constants $J_3$ and $J_1$, respectively. The coupling of a spin
$\frac{1}{2}$ with a spin 1 is intermediate. 

\begin{figure}[h]
\centerline{\psfig{file=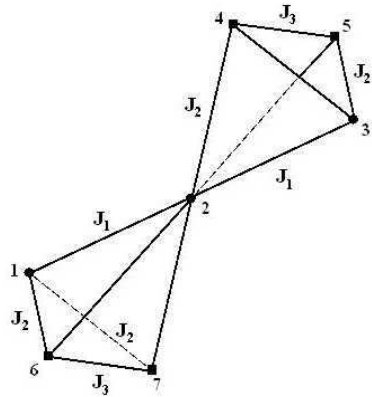,width=6cm}}
\caption[l]{Two tetrahedra of a spinel structure. Sites $\nu = 1, 2,
3$ have spin $S(\nu) = 1$ while the remaining ones have spin $\frac{1}{2}$}
\label{fig2}
\end{figure}

Consider first two corner-sharing tetrahedra (see Fig. 2) with spins ${\bf
S}(\nu)$. Let sites 1, 2, 3 have spin $S = 1$ while the remaining sites have
spin $S = \frac{1}{2}$. We introduce $\vec{\Sigma}_1 = {\bf S}(1) + {\bf
S}(2)$, $\vec{\Sigma}_2 = {\bf S}(2) + {\bf S}(3)$,  $\vec{\sigma}_1 = {\bf
S}(4) + {\bf S}(5)$,  $\vec{\sigma}_2 = {\bf S}(6) + {\bf S}(7)$ in terms of
which the interaction Hamiltonian $\sum_{\langle ij \rangle} J_{ij}{\bf
S}(i){\bf S}(j)$ is written

\begin{eqnarray}
\label{5} H_W =&& \frac{J_1}{2} \Big[\vec{\Sigma}_1^2 + \vec{\Sigma}_2^2 -8
\Big] + \frac{J_2}{2} \Big[\vec{\Sigma}_1 \vec{\sigma}_2 + \vec{\Sigma}_2
\vec{\sigma}_1\Big] \nonumber\\  
&+& \frac{J_3}{2} \bigg[\vec{\sigma}_1^2 + \vec{\sigma}_2^2 - \frac{3}{2}\bigg]
~~~.   
\end{eqnarray}

The ground state has $\sum_1 = \sum_2 = \sigma_1 = \sigma_2 = 0$. When $J_1
\gg J_2, J_3$ the low-energy excitations are given by changing
$\sigma_1$ or $\sigma_2$ from 0 to 1, i.e., they are within the spin
$\frac{1}{2}$ subsystem. They are unaffected by $J_2$, i.e., the coupling
between spin $\frac{1}{2}$ and spin 1 sites. The argument builds on the fact
that the coupling $J_1$ between the spin 1 sites is much stronger than the one
between the spin $\frac{1}{2}$ sites, i.e., $J_3$. 
It should hold also when instead of two tetrahedra the whole lattice is
considered. The excitations of spin $\frac{1}{2}$ chains and large rings are
effected by the $S = 1$ subsystem only via spin flips of the latter. However,
the excitations of the $S = 1$ subsystem have a gap and moreover from the
LDA+$U$ calculations it follows that $J_1 \simeq 8J_3$. Therefore the low-lying
excitations take place within the $S = \frac{1}{2}$ subsystem
only. Consequently $\gamma$ is given by that of uncoupled $S = \frac{1}{2}$
chains (or large rings), i.e., by (\ref{4}). Indeed, this reasoning provides
for a simple explanation of the heavy quasiparticle excitations found in
LiV$_2$O$_4$. In order to explain the experimentally observed value of $\gamma
= 0.42 J/mol \cdot K^2$ a coupling constant $J_3/k_B = 13.3 K$ is
required. This value agrees up to a factor of 2.6 with the value $J_3^{\rm
calc.}/k_B = 35 K$ found from the LDA+$U$ calculations. Furthermore, the
observed Sommerfeld-Wilson ratio of 1.7 is relatively close to the theoretical
value of $R_W = 2$ found here. Some comments are in order at that point. We
cannot expect that a LDA+$U$ calculations give precise values of spin
coupling constants, since those physical quantities are rather sensitive to a
precise description of correlations (see, e.g., \cite{ref22}) which would
require much more sophisticated computations. Also $\gamma$ as given above may
well contain contributions from phonons, i.e., $\gamma = \gamma_0 (1+\lambda)$
where $\lambda$ describes the contribution from the electron-phonon
interactions. For example, a value of $\lambda = 1$ would double the value of
$J_3$ needed for explaining $\gamma_0$ to $J_3 \approx 27 K$. But it would also
double the experimental value of $R_W$ to 3.4 since $\lambda$ does not
contribute to $\chi_{sp}$. The calculated value of $R_W = 2$ is a result for
the spin $\frac{1}{2}$ subsystem. Therefore any contribution of the spin 1
subsystem \cite{ref19} to the measured values of $\chi_{sp}$ must be subtracted
before a comparism with experiments is made. Even a small contribution of 15
$\%$ reduces $R_W$ from 3.4 to 2.7.

Experiments show no maximum in $\gamma(T) = C(T)/T$ in distinction to theory
\cite{ref18} and a broad one in $\chi_{sp}(T)$ around $T = 16 K$
\cite{ref1,ref24,ref25}. Theory predicts such a maximum at $0.6 J_3$ which
would imply a value of $J_3 \simeq 27 K$.

Although the agreement between theory and experiment is not perfect, which
comes to no surprise in view of the simplifications made, we believe to have
found a simple physical picture for the hitherto unexplained unusual
low-temperature thermodynamic properties of the material.

It is based on the proposition that the system is a superposition of
configurations with spaghetti like spin $\frac{1}{2}$ chains and rings. They
are well separated from each other since the spin 1  chains and rings
surrounding them have a large coupling constant $J_1$ and a gap in the
excitation spectrum (Haldane gap). Therefore, the low-energy excitations take
place in spin $\frac{1}{2}$ Heisenberg chains only similarly as in
Yb$_4$As$_3$. Because of the near separation of spin and charge degrees of
freedom LiV$_2$O$_4$ should not be a
Fermi liquid in the classical sense. Rather it should show features of a
Luttinger liquid because of the presence of an intrinsing chain structure even
in the cubic lattice. It would be most interesting to determine the effective
masses at the Fermi surface, e.g., by de Haas-van Alphen experiments. We
speculate that they should be much smaller than suggested by the large value of
$\gamma$. As pointed out above the tetrahedron rule is not expected to hold
rigorously. Therefore, the influence of small deviations from it remains an
interesting topic which will be investigated separately. The same holds true
for the effect of impurities and moreover for doping. Furthermore, it remains
to be seen whether similar ideas apply also to LiTi$_2$O$_4$ which is a
superconductor at low temperatures or to pyrochlore lattice systems.

\end{multicols}

\begin{center}
{\LARGE\bf *~~ *~~ *}
\end{center}

\noindent We would like to thank D. C. Johnston and H. G. von Schnering for
stimulating discussions. 

\end{document}